\documentclass[11pt,twoside]{article}

\usepackage{asp2006}
\usepackage{epsf}
\usepackage{psfig}
\usepackage{lscape}
\usepackage{graphicx}

\begin{document}
\title{Modeling He-rich subdwarfs through the Hot-Flasher scenario: First
Results} 
 
\author{Miller Bertolami M. M.$^{1,2}$ andAlthaus L. G.$^{1,2}$}
 \affil{$^1$ Facultad de Ciencias Astron\'omicas y
Geof\'{\i}sicas, UNLP, Argentina\\ $^2$ Instituto de Astrof\'{\i}sica La
Plata, UNLP-CONICET, Argentina} 
 
\begin{abstract} 

We present first results from evolutionary simulations aimed at exploring the
Hot-Flasher scenario for the formation of H-deficient subdwarf stars. The two
types of late hot flashers that lead to He-enriched surfaces, ``deep'' and
``shallow'' mixing cases, are investigated for different metallicities.

\end{abstract}

{\bf Introduction:}
Hot subluminous stars are an important population of faint blue stars which
can be roughly grouped into the cooler sdB stars and the hotter sdO
stars. Subluminous O, B stars have been identified with stars populating the
hot end of the horizontal branch of some globular and open clusters. While sdB
stars form a nearly homogeneous spectroscopical class, a large variety of
spectra is observed among the sdO stars (Stroeer et al. 2007). This variety of
spectra, and the helium enhanced surface abundances pose a challenge to
stellar evolution. As a consequence some non-canonical evolutionary scenarios
have been proposed for their formation. Among them the ``Hot Flasher
scenario'' (Castellani \& Castellani 1993, D'Cruz et al. 1996, Sweigart 1997)
offers one of the best possibilities (however, see Heber, this proceedings)
for explaining their formation, while also explaining the correlation between
the existence of blue hook stars and helium-core white swarfs (Hansen 2005,
Kalirai et al. 2007).
\bigskip

{\bf Description of the work and Results:}
We performed full evolutionary simulations of the two flavors of the
Hot-Flasher scenario, ``deep'' (DM) and ``shallow'' (SM) mixing cases (Lanz et
al. 2004), for two different metallicities (Z=0.001, 0.03). Initial values
were chosen to obtain ages, at the moment of the helium core flash, similar
to those observed in globular clusters. Initial abundances were chosen in
agreement with the evolution of galactic abundances presented by Flynn
(2004). Surface abundaces of our simulations are displayed in Table 1 and
Fig. 1 shows the log $T_{\rm eff}$-$g$ values of the sequences.
\begin{table*}[ht!]
\begin{center}
{\footnotesize

ZAMS: Mass= 0.88 M$_\odot$, X/Y/Z= 0.769/0.230/0.001. Age at Helium Flash=
12404 Myr

\begin{tabular}{c|c|c|c|c|c|c}
Final Mass [M$_\odot$]& H & He &  $^{12}$C  &  $^{13}$C  &  N   &  O  \\ \hline
0.4915 (SM) & 0.21 & 0.323 & 0.038 & $1.5 \times 10^{-6}$ & 0.0001 & 0.0014 \\
0.4910 (DM) & $1.4 \times 10^{-5}$ & 0.955 & 0.0284 & 0.0076 & 0.0086 &  $7 \times 10^{-5}$ \\
0.4815 (DM) &  $1.2 \times 10^{-6}$ & 0.965 & 0.0139 & 0.0047 & 0.0163 &  $4 \times 10^{-5}$ \\
\end{tabular}
\smallskip

ZAMS: Mass= 1.04 M$_\odot$, X/Y/Z= 0.668/0.302/0.030. Age at Helium Flash=
12863 Myr
\begin{tabular}{c|c|c|c|c|c|c}
Final Mass [M$_\odot$]& H & He &  $^{12}$C  &  $^{13}$C  &  N   &  O  \\ \hline
0.4657 (SM) &  0.512 & 0.457 & 0.00297 & 0.000121 & 0.00698 & 0.0103 \\
0.4649 (SM) &  0.026 & 0.905 & 0.0362 & $1.3 \times 10^{-5}$  & 0.00622 & 0.0019 \\
0.4633 (SM) &  0.00195 & 0.927 & 0.0371 & 0.00344 & 0.00694 & 0.00158 \\
0.4628 (DM) &  $1.3 \times 10^{-5}$ & 0.933 & 0.0424 & 0.00155 & 0.00705 & 0.00154 \\
0.4531 (DM) &  $4.5 \times 10^{-6}$ & 0.939 & 0.031 & 0.00474 & 0.01205 & 0.00198 \\

\end{tabular}
\label{tabla}
\caption{Surface abundances (by mass fractions) of the secuences calculated in
this work. The choice of the initial values of the sequences (at ZAMS) are
shown on top of each table.}}
\end{center}
\end{table*}
\begin{figure}[h!]
\begin{center}
\includegraphics[ height=112 pt] {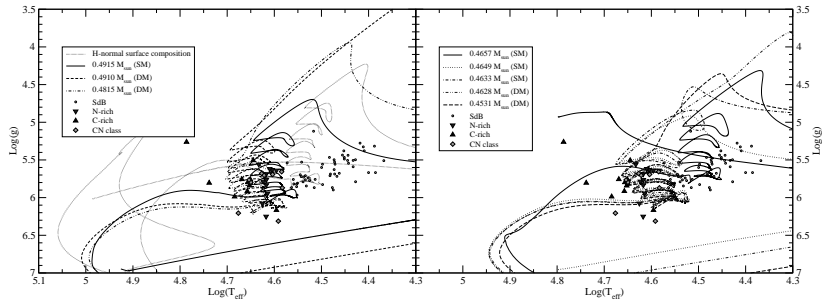}

\caption{\footnotesize Comparison of the values of $T_{\rm eff}$ and $g$ of
  our sequences for different metallicities (right panel: Z=0.03, left panel:
  Z=0.001) as compared with the values inferred by Stroeer et al. (2007).}
\end{center}
\end{figure}

We find that all our sequences display C/N$>1$ (by mass), and thus should be
considered ``carbon rich' contrary to what is inferred at some He-sdO stars,
which present both C/N$>1$ and C/N$<1$ surface abundances (Stroeer et
al. 2007). Also, consistent with previous works, we find that the range of
masses for late hot flashers is narrow and slightly dependent on mass. As
noted by Lanz et al. (2004) we also find that the proportion of shallow mixing
cases is dependent on mass, being $<5$\% at $Z=0.001$ and $\sim17$\% for
Z=0.03.

The surface temperatures and gravities of our sequences during the He-burning
phase are cooler and show a narrow range of gravity values, which are not
consistent with those inferred in He-SdO stars by Stroeer et
al. (2007). Consequently our preliminary results do not seem to support the
hot flasher origin for these stars unless strong systematics are present in the
observational inferences (Lanz et al. 2004, Heber private communication).

{\footnotesize \acknowledgements M3B thanks the organizers of the Third
Conference in Hydrogen Deficient Stars for financial assistance that allowed
him to participate in the conference.}
%
%

\end{document}